\documentclass[aip, amsmath, amssymb, reprint]{revtex4-1}
\usepackage[utf8]{inputenc}
\usepackage{physics}
\usepackage{mathrsfs}
\usepackage{amsmath}
\usepackage{amssymb}
\usepackage{mathtools}
\usepackage[margin=1in]{geometry}
\usepackage{graphicx}
\usepackage{dsfont}
\usepackage{MnSymbol}
\usepackage{fancyhdr}
\usepackage{natbib}
\usepackage{array}

\usepackage{dcolumn}
\usepackage{bm}
\usepackage[T1]{fontenc}
\usepackage{etoolbox}

\usepackage{hyperref}
\hypersetup{
  colorlinks=true,
  linkcolor=blue,
  filecolor=blue,
  urlcolor=blue,
  citecolor=blue,
}

\usepackage{ulem}

\begin{document}

\newcommand{\lbnlafil}{Applied Mathematics and Computational Research Division, Lawrence Berkeley National Laboratory, Berkeley, California 94720, USA}
\newcommand{\foundryafil}{Molecular Foundry Division, Lawrence Berkeley National Laboratory, Berkeley, California 94720, USA}
\newcommand{\lbnlmatafil}{Materials Science Division, Lawrence Berkeley National Laboratory, Berkeley, California 94720, USA}
\newcommand{\qnlafil}{Quantum Nanoelectronics Laboratory, Department of Physics, University of California, Berkeley, California 94720, USA}
\newcommand{\berkeleyphysafil}{Department of Physics, University of California, Berkeley, California 94720, USA}
\newcommand{\KavliENSI}{Kavli Energy NanoScience Institute at the University of California, Berkeley and the Lawrence Berkeley National Laboratory, Berkeley, California 94720, USA}

\title{Performance of Superconducting Resonators Suspended on SiN Membranes}
\author{Trevor Chistolini}
\thanks{Author to whom correspondence should be addressed: trevor\_chistolini@berkeley.edu}
\affiliation{\berkeleyphysafil}
\affiliation{\lbnlafil}

\author{Kyunghoon Lee}
\author{Archan Banerjee}
\affiliation{\berkeleyphysafil}
\affiliation{\lbnlmatafil}

\author{Mohammed Alghadeer}
\author{Christian Jünger}
\affiliation{\berkeleyphysafil}
\affiliation{\lbnlafil}

\author{M. Virginia P. Altoé}
\author{Chengyu Song}
\affiliation{\foundryafil}

\author{Sudi Chen}
\author{Feng Wang}
\affiliation{\berkeleyphysafil}
\affiliation{\lbnlmatafil}
\affiliation{\KavliENSI}

\author{David I. Santiago}
\affiliation{\lbnlafil}

\author{Irfan Siddiqi}
\affiliation{\berkeleyphysafil}
\affiliation{\lbnlafil}
\affiliation{\lbnlmatafil}
\affiliation{\KavliENSI}

\date{12 September 2024}

\begin{abstract}
Suspending devices on thin SiN membranes can limit their interaction with the bulk substrate and reduce parasitic capacitance to ground. While suspending devices on membranes is used in many fields including radiation detection using superconducting circuits, there has been less investigation into maximum membrane aspect ratios and achievable suspended device quality, metrics important to establish the applicable scope of the technique.
Here, we investigate these metrics by fabricating superconducting coplanar waveguide resonators entirely atop thin ($\sim$110 nm) SiN membranes, where the membrane's shortest length to thickness yields an aspect ratio of approximately $7.4 \times 10^3$. 
We compare these membrane resonators to on-substrate resonators on the same chip, finding similar internal quality factors $\sim$$10^5$ at single photon levels.
Furthermore, we confirm that these membranes do not adversely affect resonator thermalization and conduct further materials characterization. By achieving high quality superconducting circuit devices fully suspended on thin SiN membranes, our results help expand the technique's scope to potential uses including incorporating higher aspect ratio membranes for device suspension and creating larger footprint, high impedance, and high quality devices.

\end{abstract}

\maketitle

Thin silicon nitride (SiN) membranes have found use in many different applications. The membranes can be devices themselves such as high-$Q$ mechanical resonators \cite{Zwickl_SiN_high_Q, Yuan_SiN_Qi}, or they can host devices atop them for partial isolation from the substrate.
This isolation can achieve different goals - it can increase a device's sensitivity to events of interest occurring at the membrane as is popular in the radiation detector community through phonon recycling \cite{Fyhrie_responsivity_boosting, de_Visser_phonon_trapping, Bueno_KID_1um, Baselmans_MKID_100nm, Giachero_thermal_KID_2018}, limit interaction with events occurring in the bulk substrate over which it spans \cite{Lindeman_bolometers, Karatsu_MKID}, or modify a device's thermal coupling to the substrate bath \cite{Wassell_x-ray_microcalorimeter_2017, Giachero_thermal_KID_2018, Daal_microcalorimeter_update_2024, Scott_thin_SiN_thermalization_2024}. 
The membranes have also been used in electromechanics, making use of the reduction in parasitic capacitance to enable higher impedance devices, where Al coil resonators have been suspended on SiN membranes \cite{Fink_electromechanics, Fink_microwave_conversion} or on Si in a SOI platform \cite{Dieterle_electromechanics_SOI}. Moreover, SiN membranes have also been used in photonics, for purposes such as providing a transparent surface for optical access \cite{Micó_SiN_photonics, Kim_trapping_atoms_2019, Chang_microring_res_2019}.

Despite all these uses, further exploration into achievable suspended device quality on high aspect ratio membranes could drive increased progress in current implementations and potentially expand the technique to other applications or fields. Indeed, as one example, there has been comparatively little exploration of SiN membrane integration into superconducting circuits for quantum computing. However, with demonstrations of reducing correlated events in microwave kinetic inductance detector (MKID) arrays \cite{Karatsu_MKID, Day_detector_array, Mazin-review}, suggesting that such a technique could also mitigate correlated errors in quantum computing with superconducting circuits \cite{Martinis_qp, Wilen_correlated, Vepsalainen_qp_ionizing_rad, Cardani_qp_mountain, McEwen_correlated}, in addition to the membrane decreasing parasitic capacitance to ground, advantageous to qubit designs utilizing very high impedance elements such as the blochnium\cite{Pechenezhskiy_blochnium} or $0-\pi$\cite{Gyenis_0-pi}, membranes could prove highly useful to certain cases in quantum computing.

In this work, we examine the quality of devices suspended on high aspect ratio SiN membranes, after optimizing fabrication protocols that incorporate common materials of the superconducting circuits architecture of quantum computing.
We explore achievable device quality using Nb coplanar waveguide (CPW) resonators and measuring their internal quality factor ($Q_i$), which probes the material quality \cite{Gao_thesis, McRae_res, Altoe_boe}. 
While suspending devices using SiN membranes remains an active technique, we distinguish our work by our different experimental focus, materials, and design. We concentrate on achieving high $Q_i$ values across many Nb CPW resonators comparing fully suspended and non-suspended variants, and we examine those resonators using multiple methods including materials characterization to offer insight into the environment quality yielded by the membrane. 
Ultimately, our results constitute an advancement in the fabrication of high aspect ratio SiN membranes for device suspension, help to expand their scope in applications using very large footprint or high impedance devices, and validate their compatibility with uses requiring low loss environments, such as quantum computing with superconducting circuits.

\begin{figure}
    \centering
    \includegraphics[scale=1]{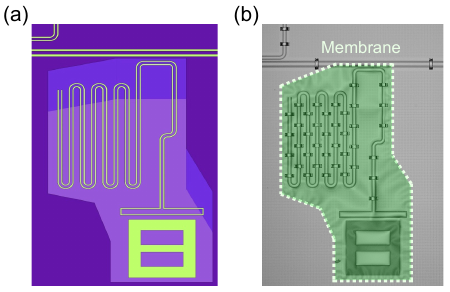}
    \caption{Suspended CPW resonators. (a) Design concept of the suspended resonator, displaying the transparent membrane where all bulk Si beneath the resonator has been entirely etched away. (b) Optical image of the physical device, primarily displaying a CPW atop a SiN membrane, where the membrane is false colored in green. At the top left, a section of an ``on-substrate" resonator is visible, and the central transmission line lies between the resonators. Aluminum airbridges span across all signal traces.}
    \label{fig:intro}
\end{figure}

The CPW's were patterned on 10 mm $\times$ 10 mm Si chips, each consisting of eight resonators - four directly on the bulk substrate (on-substrate condition) and four on thin SiN membranes spanning the substrate (membrane condition). This examination of a CPW fabricated entirely on a SiN membrane offers insight into the quality factor and characteristics of such devices. All resonators were coupled to a central transmission line and measured in reflection, with a design image and optical image of a suspended resonator shown in Fig. \ref{fig:intro}(a) and (b). Due to the membrane's fragility, airbridges were necessary to galvanically stitch together the ground plane to reduce parasitic slotline modes, as opposed to using wirebonds \cite{Hornibrook_CPW_dissipative, Chen_airbridges, Sun_airbridges}. 
We present the results from two samples principally characterized in an adiabatic demagnetization refrigerator (ADR) with a base temperature of 100 mK. A wiring diagram of this measurement setup is included in the Supplementary Information.

\begin{figure}
    \centering
    \includegraphics[scale=1]{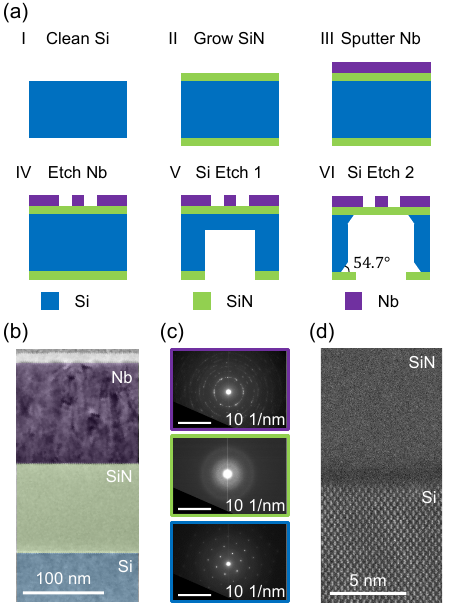}
    \caption{Fabrication procedure and materials characterization. (a) Overall steps in fabricating CPW resonators atop SiN membranes. (b) A cross-sectional TEM image exhibiting the distinct layers of the system, false colored to match panel (a). (c) Electron diffraction patterns from the Nb, SiN, and Si in descending order. These measurements indicate the Nb to be polycrystalline and the SiN to be amorphous. The Si single crystal substrate is seen at the [011] zone axis. (d) An atomic resolution STEM-HAADF image of the SiN and Si, revealing an interface layer of $\sim$1 nm.}
    \label{fig:fab}
\end{figure}

\begin{figure*}[t!]
    \centering
    \includegraphics[scale=1]{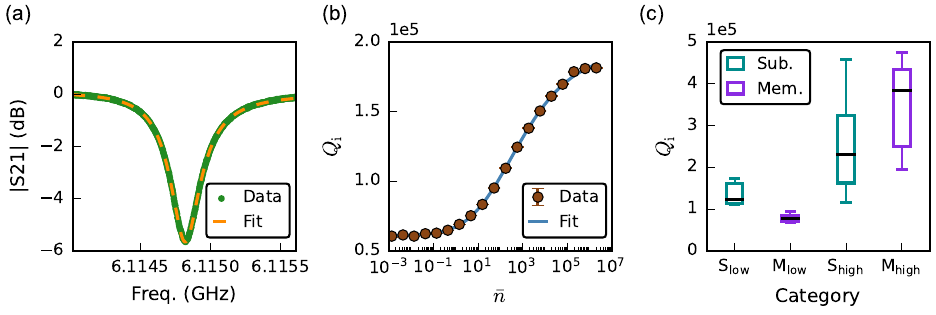}
    \caption{Internal quality factor measurements. (a) A representative scan across a single resonator, revealing the dip in the amplitude response on resonance. (b) $Q_i$ versus $\bar{n}$ of the resonator shown in (a), extracted by fitting the resonator response while increasing the drive power. A fit using Eq. (\ref{eq:internal_loss}) is overlaid. (c) Summarizing $Q_i$ statistics of all measured on-substrate (S, teal) and membrane (M, purple) resonators, in low and high power regimes.}
    \label{fig:Qi}
\end{figure*}

The device fabrication combined methods from superconducting circuits for quantum computing with those for fabricating SiN membranes, with the major steps illustrated in Fig. \ref{fig:fab}(a), and further details are presented in the Supplementary Information. The CPW resonators were etched from a $\sim$125 nm thick Nb layer atop a $\sim$110 nm SiN layer, which was grown on a high-resistivity 675 $\mu$m thick Si wafer. The SiN was deposited using low pressure chemical vapor deposition (LPCVD), and silicon-rich $\text{SiN}_x$ was chosen instead of $\text{Si}_3\text{N}_4$ due to its lower dielectric loss \cite{Paik_SiN_loss}. The thickness of each layer was measured using cross-sectional transmission electron microscopy (TEM), shown in Fig. \ref{fig:fab}(b). Further details on the materials characterization can be found in the Supplementary Information.
The fabrication posed various challenges due to the fragility of the membranes, especially at the dimensions of this work. The ratio between the shortest (longest) span of the etched region and the SiN membrane thickness was approximately $0.7\times 10^4$ ($2.2\times 10^4$). 
These aspect ratios are encouraging advances over those commonly used in membranes for device suspension, and the membranes were robust to repeated thermal cycling. While fabrication steps in our procedure have been used in past works, we found a combination of select less common steps necessary for our process. For one, a two-step dry then wet Si etch was critical to accurately achieve target membrane dimensions, likely due to our relatively thick substrate, where the initial dry etch removed most of the Si \cite{Kim_trapping_atoms_2019}. Furthermore, a wafer holder with an O-ring to isolate the wafer topside provided reliability during KOH etching \cite{Fink_electromechanics}. Finally, the presence of fragile airbridges on our devices demanded metal liftoff to occur after the final Si etching, which required optimization of the liftoff procedure followed by critical point drying to preserve the membranes.

We used a variety of techniques to characterize the resonators in the frequency domain, with the prime metric of interest being their internal quality factor. We collected the resonator scattering parameters (S21) using a vector network analyzer (VNA), which were then fit to a reflection model using methods described in past work \cite{Altoe_boe}, as shown in Fig. \ref{fig:Qi}(a). From this, we extracted the resonator $Q_i$, or its inverse the internal loss $\delta_i = 1/Q_i$.

One significant loss source for resonators are two-level systems (TLS), whose effects are expected to saturate at high powers \cite{Pappas_TLS}. Thus, if TLS are a dominant loss source, $Q_i$ is expected to increase with increasing power before saturating, and the difference between $Q_i$ at low and high powers distinguishes the relative loss contribution from TLS versus other sources (note, in certain devices such as Al resonators at low temperature, $Q_i$ can instead decrease with increasing drive power as more quasiparticles are generated \cite{Visser_Al_CPW_qp}). The relevant $Q_i$ values can be obtained by probing the resonator with a varying power to obtain $Q_i$ versus average photon number ($\bar{n}$), which can be fit, as in Fig. \ref{fig:Qi}(b), by

\begin{equation}\label{eq:internal_loss}
    \delta_{i}(\bar{n}, T) = \delta_{\text{TLS}}^0 \frac{\tanh{\left( \frac{hf}{2k_B T} \right)}}{\sqrt{1+(\bar{n}/n_c)^\beta}} + \delta_{\text{other}}(T).
\end{equation}
Here, $\delta_{\text{TLS}}^0$ is the low power TLS loss, $\beta \leq 1$ is a filling factor, $n_c$ is a critical photon number, and $\delta_{\text{other}}$ is a power-independent loss term that $\delta_i$ will saturate to at high drive powers \cite{Altoe_boe}. 

The on-substrate versus membrane resonators were compared by their low and high power $Q_i$ values, as shown in Fig. \ref{fig:Qi}(c), with the individual resonator plots included in the Supplementary Information. With reference to Eq. (\ref{eq:internal_loss}), the low power values correspond to $Q_i = [\delta_i(\bar{n}=1, T=\text{100 mK})]^{-1}$, and the high power values correspond to $Q_i = [\delta_\text{other}(T = \text{100 mK})]^{-1}$. 
At low power, the membrane (on-substrate) resonators had a median $Q_i$ of $0.8 \times 10^5$ ($1.2 \times 10^5$); meanwhile, at high power, the membrane (on-substrate) resonators had a median $Q_i$ of $3.8 \times 10^5$ ($2.3 \times 10^5$). However, as seen in panel (c), the high power results were comparable to each other within their spreads. 

One possible explanation of this difference at low power is a greater amount of TLS affecting the membrane resonators. Those resonator's fields occupy both the Nb metal-air interface and the bottom SiN-air interface, in contrast to the on-substrate resonators that do not encounter the latter air interface but instead a typically cleaner SiN-Si interface \cite{Wenner_res_sim, Woods_res}. This SiN-Si interface was more closely examined to confirm its relatively low levels of oxides, as shown within the Supplementary Information.

All resonators had aluminum (Al) airbridges spanning over the CPW signal trace. As the device temperature was increased closer to the superconducting transition temperature of Al ($T_c \approx 1.2\text{ K}$) \cite{Caplan_Al_TC}, we found systematic changes in the resonator frequency and $Q_i$ that can be well fit using a cavity-perturbation model \cite{Gruner_cavity_perturbation_1993}, based on the Mattis-Bardeen theory~\cite{Mattis_Bardeen_1958} (see Supplementary Information for more details). This indicates that the temperature evolutions were mainly associated with the changing Al surface impedance, instead of factors such as fabrication processing.

\begin{figure}
    \centering
    \includegraphics[scale=1]{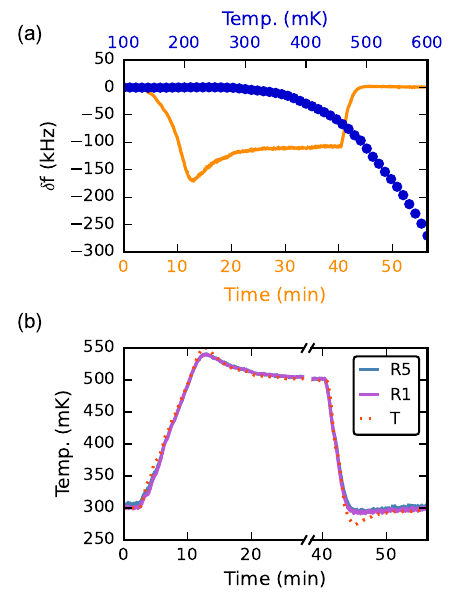}
    \caption{Temperature thermalization scans. (a) Twin axes plot. In blue, the frequency shift versus temperature for a representative resonator is shown, exhibiting a decreasing frequency with increasing temperature. In orange, the frequency response of the same resonator is shown as the ADR temperature is swept from 300 - 500 - 300 mK. (b) The temperature response of two resonators as the ADR temperature (dotted red line) is swept, extracted using calibration scans as shown in (a) in blue to convert between frequency and temperature.}
    \label{fig:temp}
\end{figure}

Leveraging this effect, we investigated whether the resonators on bulk substrate significantly differ in their thermalization rate compared to those on membrane. 
To do so, we first increased the cryostat's temperature in 10 mK steps, and each resonator's scattering parameters were measured at high power ($\bar{n} \sim 10^4$), chosen to ensure later scans could be taken quickly without averaging. From the resulting frequency versus temperature, as shown in blue in Fig. \ref{fig:temp}(a), we fit the relationship in the temperature regime of interest to a monotonic function, thus enabling the conversion between frequency and temperature. After these calibration measurements, the ADR's base plate was swept up from 300 mK to 500 mK, held, and then swept back down. Throughout this process, the resonator response was monitored, with its fitted frequency shown in orange in Fig. \ref{fig:temp}(a). For each experiment, two resonators were monitored in an interleaved fashion to achieve a faster repetition rate. This experiment was repeated, where one resonator in the pair remained the same, while the other resonator changed. Although this did not allow for a comparison among all resonators from a single sweep, a thermalization difference between on-substrate versus membrane resonators would still be apparent in these pairwise sweeps. 

Using the calibration scans to convert from frequency to temperature, we compared the temperatures versus time of each resonator in a pair, as in Fig. \ref{fig:temp}(b), where the thermometer reading of the cryostat is also displayed. For each of these experiments, we identified four time intervals where any thermalization differences between resonators should be more apparent. 
These correspond, in order, to when the ADR temperature initially increased from the low set point, when it maximized after reaching the high set point, when it initially began to decrease back to the low set point, and when it minimized after reaching the low set point again. 
If one resonator thermalized faster than the other, its temperature would change sooner in the first and third intervals, while it would achieve its maximum (minimum) temperature sooner in the second (fourth) interval. 

For all resonator pairs, no systematic difference in the thermalization rates was observed. We note that in these experiments sweeping between 300 mK - 500 mK, how the resonators thermalize to the bath may differ to that at sub-100 mK temperatures, relevant to certain applications. Furthermore, these indistinguishable thermalization rates may seem unexpected considering how SiN membranes are used for thermal isolation from the substrate\cite{Wassell_x-ray_microcalorimeter_2017, Giachero_thermal_KID_2018, Daal_microcalorimeter_update_2024, Scott_thin_SiN_thermalization_2024}, related to SiN's low thermal conductivity\cite{Jianqiang_Si_rich_SiN_thermal_2014, Ftouni_SiN_thermal_stress_2015} However, in contrast to work where the suspended device may not have metal connecting it to the region on substrate, the Nb ground plane of our chips does connect the suspended CPW's to regions over the bulk substrate, which may mediate more effective thermalization. Further details on this thermalization experiment are presented in the Supplementary Information.

In this work, we fabricated and characterized suspended Nb CPW resonators on thin SiN membranes, finding high-$Q$ superconducting circuits compatible with such a technique. The membrane resonators exhibited similar thermalization rates to their on-substrate counterparts. Furthermore, no difference was observed in their temporal frequency stabilities, with further details presented in the Supplementary Information. The on-substrate resonators also enabled a benchmark of our general fabrication quality, which was further investigated through materials characterization measurements. Indeed, the comparable resonator internal quality factors $\sim$$10^5$ suggests that suspending resonators does not limit their $Q_i$ to our measured values, and our device quality was instead currently limited by the general fabrication. 
While these quality factors remain below state-of-the-art levels for conventional on-substrate CPW's \cite{Megrant_high_Q, Altoe_boe}, increasing membrane thickness for robustness to enable more aggressive cleaning and address the buckling observed in Fig. \ref{fig:intro}(b), initially annealing the SiN \cite{Mittal_annealing_SiN}, or using the ``SC-2" cleaning method to remove potassium ions following the KOH etch \cite{Bueno_KID_1um, Baselmans_MKID_100nm} could yield improvements. 

Looking forward, our results have implications to a variety of fields that already use SiN membranes in addition to others which may benefit from their incorporation. By achieving these high aspect ratio membranes for device suspension, our fabrication results expand upon previous implementations and could guide further development in suspending larger footprint devices or devices on even thinner membranes, which could increase kinetic inductance detector (KID) efficiency \cite{Fyhrie_responsivity_boosting} or improve thermal isolation \cite{Scott_thin_SiN_thermalization_2024}. At the same time, our results suggest that suspended devices can still be well thermalized to the bath provided there exist other connections such as the Nb metal, so device suspension can be applied to uses needing greater thermalization. 
Due to the reduction in parasitic capacitance to ground caused by the membrane, relevant devices to suspend include larger high impedance elements such as superinductors used in certain qubit designs \cite{Pechenezhskiy_blochnium, Gyenis_0-pi}. Indeed, this property of membranes has been utilized to create high impedance elements \cite{Fink_electromechanics, Peruzzo_geometric_superinductor_SOI_2020}.
Furthermore, by achieving suspended devices of sufficiently high $Q_i$ with results indicating they could be further improved, SiN membranes can be applied more confidently toward applications requiring low-loss environments, such as suspending qubits, which may limit correlated errors across a chip in quantum computing. Although the reduction in ground capacitance necessitates modifications to typical qubit designs, such modifications can be achieved, as described in the Supplementary Information.
\newline

See the Supplementary Information for further information referenced within the text.
\newline

We thank Bingcheng Qing and Zahra Pedramrazi for assistance with cryogenic setups. We would also like to acknowledge the UC Berkeley Marvell Nanofabrication Laboratory, where sample fabrication was principally performed.

T.C. acknowledges support from the National Science Foundation Graduate Research Fellowship Program (NSF GRFP) under Grant No. DGE 1752814 and DGE 2146752. This material, including work performed at the Molecular Foundry, was funded in part by the U.S. Department of Energy, Office of Science, Office of Basic Energy Sciences, Materials Sciences and Engineering Division "High Coherence Multilayer Superconducting Structures for Large Scale Qubit Integration and Photonic Transduction program (QIS-LBNL)" and by the U.S. Department of Energy, Office of Science, Office of Advanced Scientific Computing Research Quantum Testbed Program under contract DE-AC02-05CH11231.

\section*{Author Declarations}
\subsection*{Conflict of Interest}
The authors have no conflicts to disclose.

\subsection*{Author Contributions}
\noindent
\textbf{Trevor Chistolini:} Conceptualization (equal); Data curation (lead); Formal analysis (lead); Investigation (lead); Methodology (lead); Project administration (lead); Software (lead); Supervision (equal); Validation (lead); Visualization (lead); Writing -  original draft (lead); Writing - review and editing (equal). 
\textbf{Kyunghoon Lee:} Conceptualization (equal); Investigation (supporting); Methodology (supporting); Supervision (equal); Writing - review and editing (equal). 
\textbf{Archan Banerjee:} Conceptualization (equal); Investigation (supporting); Methodology (supporting); Supervision (equal); Writing - review and editing (equal). 
\textbf{Mohammed Alghadeer:} Investigation (supporting); Software (supporting); Writing - review and editing (equal). 
\textbf{Christian Jünger:} Visualization (supporting); Writing - review and editing (equal). 
\textbf{M. Virginia P. Altoé:} Data curation (supporting); Formal analysis (supporting); Investigation (supporting); Supervision (supporting); Visualization (supporting); Writing -  original draft (supporting); Writing - review and editing (equal).
\textbf{Chengyu Song:} Data curation (supporting); Investigation (supporting); Writing - review and editing (equal). 
\textbf{Sudi Chen:} Formal analysis (supporting); Methodology (supporting); Software (supporting); Supervision (supporting); Visualization (supporting); Writing -  original draft (supporting); Writing - review and editing (equal). 
\textbf{Feng Wang:} Funding acquisition (supporting); Project administration (supporting); Writing - review and editing (equal). 
\textbf{David I. Santiago:} Funding acquisition (supporting); Project administration (supporting); Writing - review and editing (equal). 
\textbf{Irfan Siddiqi:} Funding acquisition (lead); Project administration (supporting); Resources (lead); Writing - review and editing (equal).

\section*{Data Availability}
The data that support the findings of this study are available from the corresponding author upon reasonable request.

\cleardoublepage
\onecolumngrid
\newcommand{\beginsupplement}{
  \setcounter{equation}{0}
  \setcounter{figure}{0} 
  \setcounter{table}{0}
  \renewcommand{\theequation}{S\arabic{equation}}
  \renewcommand{\thefigure}{S\arabic{figure}}
  \renewcommand{\thetable}{S\arabic{table}} 
}

\section*{Supplementary Information}
\beginsupplement

\section{ADR Setup}\label{app:ADR_Wiring}
\begin{figure}[h]
    \centering
    \includegraphics[scale=1]{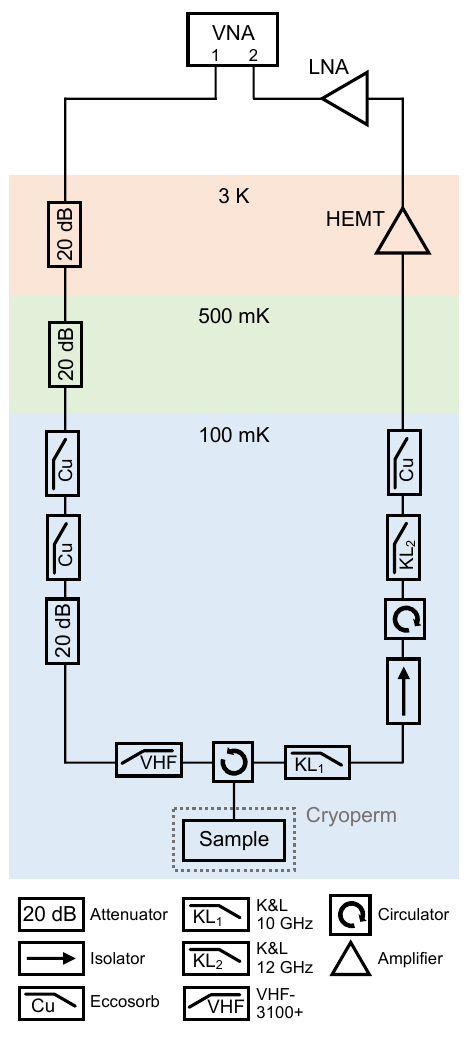}
    \caption{Diagram of the ADR cryostat and legend of its components, including attenuators, isolators, lowpass and highpass filters where their passbands are specified, circulators, and amplifiers.}
    \label{fig:ADR}
\end{figure}

Most characterization of devices was performed in an HPD Rainier 103 adiabatic demagnetization refrigerator (ADR) at a base temperature of 100 mK. The samples were measured in reflection using a central bus to interact with all resonators, minimizing the needed number of lines. A VNA was used for all measurements, and the ADR temperature could be reliably modified to perform temperature sweeps. Further details on the microwave setup are presented in Fig. \ref{fig:ADR}, similar to past work where resonator characterization was performed \cite{Altoe_boe}.

\section{Fabrication Procedure}\label{app:Fab_procedure}
While the main fabrication steps were outlined in the main text, we provide more extensive details here. To begin, a 6 inch, 675 $\mu$m thick high-resistivity ($\geq 10,000$ ohm-cm) Si wafer was cleaned using a piranha etch and a buffered-oxide etch (BOE). A 110 nm layer of Si-rich $\text{SiN}_x$ was then grown on both sides of the wafer \cite{Paik_SiN_loss}. Following this, 125 nm of Nb was sputtered on one side. A conventional optical lithography procedure was performed to write and develop the ground plane features. The Nb was then etched to yield these completed features, which were protected before moving on to write the etch windows on the opposite side of the wafer. Once developed, these windows were used to selectively etch through the SiN layer ($\text{C}_4\text{F}_8$ based) and Si wafer ($\text{C}_4\text{F}_8-\text{SF}_6-\text{O}_2$ based) beneath the resonators using consecutive anisotropic, deep reactive ion etching (DRIE) following the Bosch process, with a target total etching depth of 600 $\mu$m. 
Due to the fragility of the membranes but necessity of galvanically stitching together the ground plane, airbridges were written using a single-step grayscale process. 
After development and ion milling to ensure galvanic contact, Al was evaporated to yield the airbridges. 

Liftoff following this evaporation was delayed due to the airbridge fragility; instead, the wafer topside was protected with a dedicated photoresist before etching the remaining Si with KOH, which self-terminates at the SiN layer. This overall two-part etch series of a dry then wet etch was challenging to optimize. Use of only KOH instead of starting with a dry etch was unsuccessful due to the etch spreading too wide while etching through the Si wafer, rendering it challenging to achieve target membrane dimensions. We found that beginning with the dry, anisotropic etch was necessary. Another component to optimize revolved around how KOH also etches Nb. Consequently, any exposed feature or side of the Nb layer would lead to a wafer-scale failure during the wet etch. Thus, in addition to the protective layer, a clamping tool with an O-ring was used to only expose the etch windows to the KOH. 

With the membranes now formed but the wafer still whole, individual dies were cleaved by hand, following cleaving lines formed earlier. Using a dicing saw would be too vigorous and break the membranes. Once an individual die was obtained, liftoff was performed using room temperature acetone overnight, for around 12 hours. This differs from other liftoff techniques using hot (67$^\circ$C) acetone or hot (80$^\circ$C) or even room temperature Microposit Remover 1165, which were all too aggressive for the membranes \cite{Kreikebaum_JJ_uniformity}. Finally, the dies were transferred to IPA, where the solvent was removed using a critical point dryer (CPD). Again, removal of solvent by other means such as $\text{N}_2$ blow drying, spin drying, or a hot plate broke the membranes. After this extensive optimization, the ultimate yield of a chip with all four membranes intact was approximately $50\%$, with the final CPD step having the lowest success rate. However, utilizing a compatible sample holder for these $10 \text{ mm} \times 10 \text{ mm}$ chips could improve yield by keeping the chip positions more fixed during the CPD process.

\section{Materials Characterization}\label{app:Mat_characterization}
\begin{figure}[h]
    \centering
    \includegraphics[scale=1]{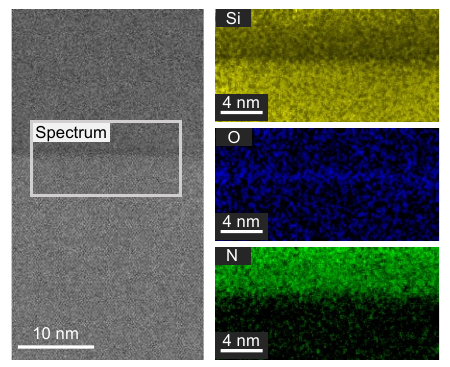}
    \caption{Elemental mapping extracted from EELS characterization of the Si-SiN interface, showing a clean interface with little oxygen contamination. The Si and N elemental maps are also displayed.}
    \label{fig:Si_interface_EELS}
\end{figure}

Materials characterization measurements were performed to examine the quality of the devices, similar to past work \cite{Altoe_boe}. In preparation for TEM imaging, cross-section lamellae were prepared by FIB machining and extracted from the resonator chips. They were prepared at the Surface Analysis Laboratory, UTAH NANOFAB using a FEI Helios Nanolab 650 dual-beam at the University of Utah. Once the samples were received back to LBNL, TEM images and EDS elemental maps were acquired at 200 kV using a JEOL 2100-F field emission scanning transmission electron microscope (STEM), which was equipped with an Oxford high-solid-angle silicon drift detector (SDD) x-ray energy dispersive spectrometer. For EDS spectral imaging, a 1-nm-diameter electron-beam probe was used. For finer resolution measurements, STEM-EELS was performed at 300 kV using the TEAM I double-aberration-corrected STEM equipped with a high-resolution Continuum Gatan imaging filter (GIF) spectrometer and a 4K $\times$ 4K Gatan K3 direct electron detector. EELS spectral imaging maps were acquired using an electron-beam probe of a 0.1 nm diameter with 1-eV elastic-peak energy-resolution. A measurement of this chemical mapping is displayed in Fig. \ref{fig:Si_interface_EELS}, showing the clean interface between the Si and SiN layers.

The resonator characterization is also notable in light of the Nb being grown on amorphous SiN, in contrast to the more standard case of growth on single crystalline Si. The Nb film on SiN exhibits a smaller grain size with random crystallographic orientation as compared to the larger grain sizes and preferred crystallographic orientation with apparent epitaxial Nb-grain nucleation observed in Nb on Si \cite{Altoe_boe}. These distinctions raise further avenues of exploration concerning Nb growth on amorphous substrates and how the differences in grain size and orientation could affect its material properties and chemistry. 

\section{Resonator $Q_i$ Details}\label{app:Res_Qi_more}
\begin{figure}
    \centering
    \includegraphics[scale=1]{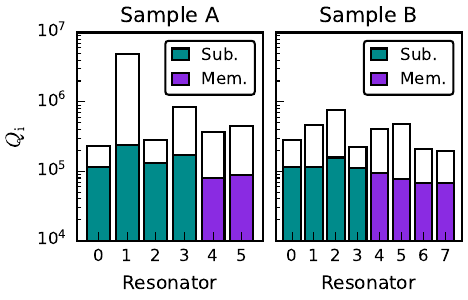}
    \caption{Internal quality factor plots for each resonator. The low and high power $Q_i$ values are presented for each resonator from Samples A and B, distinguished by the on-substrate (teal) versus membrane (purple) types. Solid colored sections indicate the low power $Q_i$, while black outlines indicate the corresponding high power values.}
    \label{fig:Qi_bar_plots}
\end{figure}

Here, we provide the measured $Q_i$ values for each resonator of the two samples, measured at low power (single photon level) and high power, as shown in Fig. \ref{fig:Qi_bar_plots}. Each chip had four resonators of each type on it. Although all resonators were found on Sample B, instead six were observed on Sample A.

Regarding the fitting routine of the resonator response versus power, in light of more noisy data at low power values, all fits were restricted to the interval above $\bar{n} \sim 10^{-3}$. 
Points were defined as outliers, likely due to noise at low powers or abnormal effects at high powers as also seen in other work \cite{Crowley_Ta_res, Drimmer_Nb_res_loss}, and removed from the fitting procedure starting at if: (1) any low power point had a $\delta_i$ that decreased by more than $10\%$ relative to its neighboring next point, and (2) any high power point had a $\delta_i$ that increased by more than $10\%$ relative to its neighboring previous point.

\section{Airbridge Effect Modelling}\label{app:Airbridge_fit}

\begin{figure}
    \centering
    \includegraphics[scale=1]{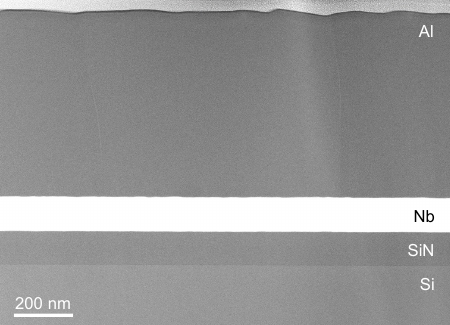}
    \caption{A STEM dark field image of the device, including the Al airbridge where it makes contact with the Nb.}
    \label{fig:Al_TEM}
\end{figure}

\begin{figure}
    \centering
    \includegraphics[scale=1]{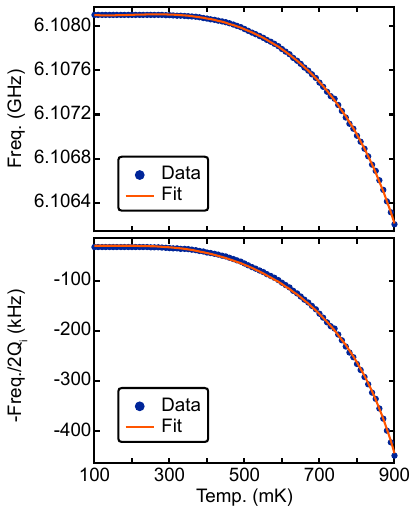}
    \caption{Temperature evolutions of resonator frequency and internal quality factor. $f$ (top) and $-f/2Q_i$ (bottom) extracted from frequency-domain measurements are plotted in blue circles versus temperature. These data are simultaneously fit to a model that includes both Al airbridge and TLS effects. The fitted curves are plotted in orange and show good agreement with the data.}
    \label{fig:airbridge_effects}
\end{figure}

The observed small relative change of frequency ($f$) with temperature ($T$) suggests that $T$-dependent effects can be considered as perturbations to the resonant mode. In this cavity-perturbation regime~\cite{Gruner_cavity_perturbation_1993}, the $T$ evolution of the complex resonant frequency $\hat{f}\coloneqq f - \frac{if}{2Q_i}$ follows
\begin{equation}\label{eq:cmplx_frequency_shift}
    \hat{f}(T)= \hat{f}_r - i g \hat{Z}_{\text{s}}(T) + \delta\hat{f}_{\text{TLS}}(T),
\end{equation}
where $\hat{f}_r$ is a complex constant, $\hat{Z}_{\text{s}}$ is the surface impedance of the Al airbridges, $g$ is a real proportionality constant, and $\delta\hat{f}_{\text{TLS}}$ is the TLS contribution to the complex frequency shift. We ignore the surface impedance change of Nb because its $T_c$ is much higher than the $T$ range studied here.

We use the following expression for the TLS contribution~\cite{Gao_thesis,Pappas_TLS}:
\begin{equation}
\label{eq:TLS_Tdep}
   \delta\hat{f}_{\text{TLS}} = \frac{\delta_{\text{TLS}}^0}{\pi}\left[\text{Re}\Psi\left(\frac{1}{2} - \frac{\hbar\omega}{2i\pi k_B T}\right)- \ln\frac{\hbar\omega}{2\pi k_B T}\right],
\end{equation}
where $\omega$ is the angular frequency, $k_\text{B}$ is the Boltzmann constant, and $\Psi$ is the complex digamma function. Because the data were collected in the high-power regime where $\bar{n}\gg n_c$, here we have omitted the imaginary part of $\delta\hat{f}_{\text{TLS}}$, which is suppressed by a factor of $\sqrt{1+(\bar{n}/n_c)^\beta} \gg 1$ in comparison to its real part. We also set $\omega$ to the measured resonant frequency at 100 mK multiplied by $2\pi$, thanks to the minute relative change of $f$.

For the airbridge contribution, as the penetration depth in Al is much smaller than the coherence length~\cite{Miller_1960} and the airbridge thickness ($\sim$615 nm as measured by TEM at the pads in contact with Nb, see Fig.~\ref{fig:Al_TEM}), we calculate $\hat{Z}_{\text{s}} = R_s + i X_s$ using
\begin{gather}
    R_s= \left(\frac{\mu_0\omega}{2}\frac{\sqrt{\sigma_1^2+\sigma_2^2}-\sigma_2}{\sigma_1^2+\sigma_2^2}\right)^{1/2},\\
    X_s= -\left(\frac{\mu_0\omega}{2}\frac{\sqrt{\sigma_1^2+\sigma_2^2}+\sigma_2}{\sigma_1^2+\sigma_2^2}\right)^{1/2},
\end{gather}
and according to the Mattis–Bardeen theory~\cite{Mattis_Bardeen_1958},
\begin{equation}
\label{eq:MB1}
    \sigma_1(\omega,T) = \frac{2\sigma_N}{\hbar \omega}\cdot
    \int_{\Delta_{T}}^{\infty}\frac{[f_T(E)-f_T(E+\hbar\omega)](E^2+\Delta_T^2+\hbar\omega E)}{\sqrt{(E^2-\Delta_T^2)[(E+\hbar\omega)^2-\Delta_T^2]}}dE,
\end{equation}
\begin{equation}
\label{eq:MB2}
    \sigma_2(\omega,T) = \frac{\sigma_N}{\hbar \omega}\cdot
    \int_{\Delta_T-\hbar\omega}^{\Delta_T}\frac{[1-2f_T(E+\hbar\omega)](E^2+\Delta_T^2+\hbar\omega E)}{\sqrt{(\Delta_T^2-E^2)[(E+\hbar\omega)^2-\Delta_T^2]}}dE.
\end{equation}
Here $\sigma_N$ is the conductivity in the normal state, $\mu_0$ is the vacuum permeability, $f_T$ is the Fermi function at $T$, and the superconducting gap $\Delta_T$ is approximated as $\Delta_0 \tanh\left({1.74 \sqrt{(T_c - T)/T}}\right)$ with $\Delta_0$ representing the zero-temperature gap size. Eqs.~(\ref{eq:MB1})~and~(\ref{eq:MB2}) are valid for $\hbar\omega<2\Delta_T$, which always holds in the temperature and frequency range of our measurements.

As shown in Fig.~\ref{fig:airbridge_effects}, we fit the real and imaginary parts of $\hat{f}$ simultaneously using Eqs.~(\ref{eq:cmplx_frequency_shift})-(\ref{eq:MB2}). The independent fitting parameters are $\hat{f}_{r}$, $g/\sqrt{\sigma_N}$, $\delta_{\text{TLS}}^0$, $T_c$, and $2\Delta_0/k_B T_c$. The integrals in Eqs.~(\ref{eq:MB1})~and~(\ref{eq:MB2}) are evaluated numerically using code modified from Ref.~\cite{Aude_2010}. The fit well reproduces the overall temperature dependence, with mild overestimation of $\delta_{\text{TLS}}^0$ hardly visible on the plotted frequency scale. The surface impedance term is found to dominate the $T$ response, and the extracted $T_c = 1.14(1)\text{ K}$ and $2\Delta_0/k_B T_c = 3.31(3)$ are in reasonable agreement with literature values for Al~\cite{Miller_1960,Caplan_Al_TC}. These results indicate that the temperature-dependent effects can mainly be attributed to the Al airbridges.

Notably, the airbridge impact on the resonator $Q_i$ and frequency offers a temperature bound beyond which Al airbridges should no longer be used, with $Q_i$ values decreasing to half their maximum at around $460$ mK. This property is important to keep in mind as efforts continue to explore operating superconducting circuits at higher temperatures \cite{Anferov_Nb_trilayer, Anferov_200mK_qubit}.

\section{Thermalization Rate Comparison Analysis}\label{app:Thermalization_comparison}
\begin{figure*}
    \centering
    \includegraphics[scale=1]{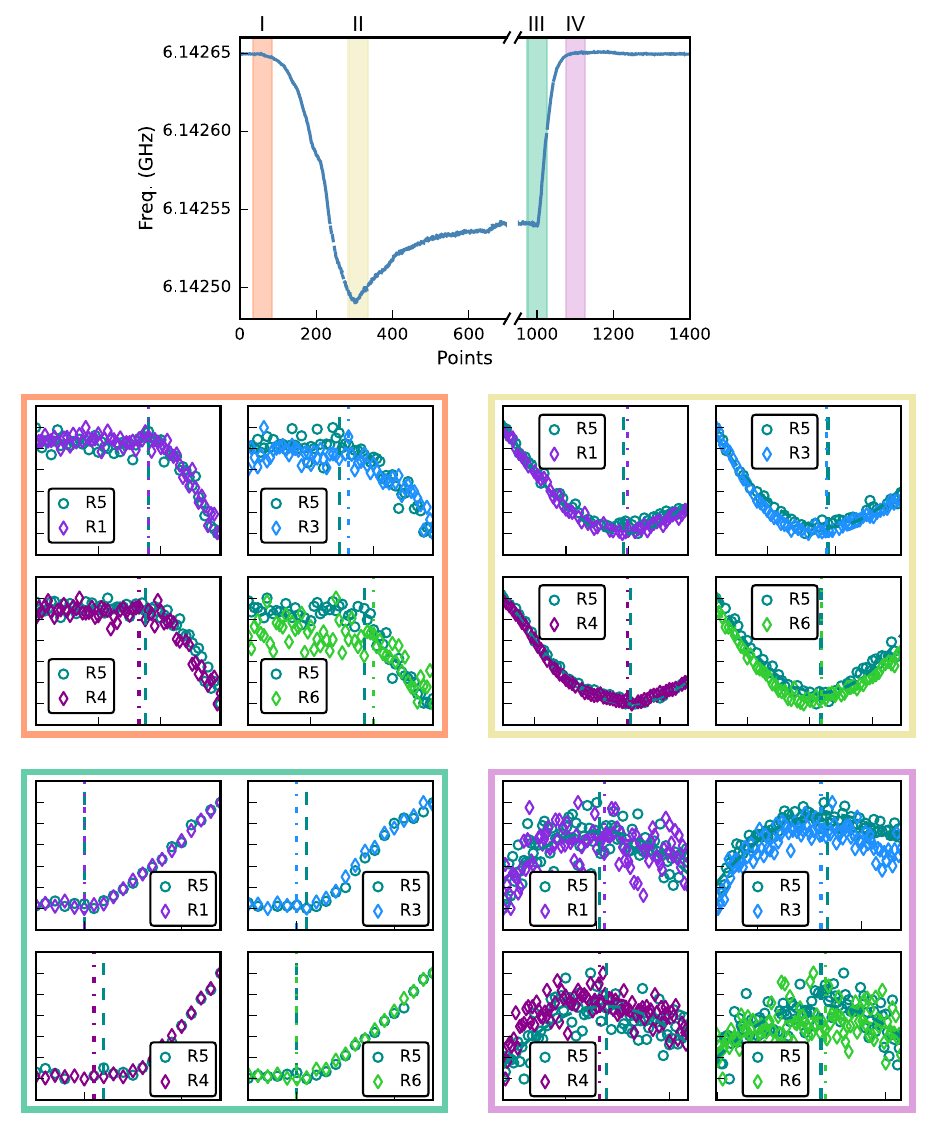}
    \caption{Comparison of temperature thermalization rates between the resonators. At the top is shown a resonator's frequency over time, where the x-axis is denoted by the recorded point, and the intervals are shaded corresponding to the bottom panels. The region labels of each interval are indicated above. In each below panel, the pairwise scan results are shown for the respective region. The vertical lines indicate the point for comparison between the resonators, where the resonator's frequency is taken to initially decrease, minimize, increase, and maximize in regions I, II, III, and IV, respectively. For all these cases, an identified line farther to the left indicates a faster thermalization rate.}
    \label{fig:temp_fit}
\end{figure*}

Here, we present further information on the comparisons between resonator thermalization rates. For each region of interest, shaded in different colors in Fig. \ref{fig:temp_fit}, the two resonator frequencies versus recorded point were examined more closely to determine any significant differences. R1 and R4 are on SiN membranes, while R3, R5, and R6 are on-substrate.

In region I (III), resonator thermalization differences would result in one resonator's frequency changing sooner than the other. To identify the point when the resonator's frequency started to change, with noise in mind, a moving average using a window of $5$ points was run across the data. The difference was taken between each window's mean and the prior window's, and the resonator's initial point of change was identified as the point when all further window differences were negative (positive), meaning that the moving averages continually became lower (higher). In region II (IV), the resonator frequency versus point was fit to an asymmetric parabola, and the minimum (maximum) points were compared. The difference in these corresponding points for each pair of resonators is presented in Table \ref{table:res_therm}.

\begingroup
\setlength{\tabcolsep}{6pt}
\renewcommand{\arraystretch}{1.2}
\begin{table}
\centering
\begin{tabular}{|c|c|c|c|c|}
 \hline
 Res. Pair & Reg. I & Reg. II & Reg. III & Reg. IV \\
 \hline 
 1-5 & 0 & 1 & 0 & 3 \\
 \hline 
 3-5 & 3 & 0 & -1 & -3 \\
 \hline 
 4-5 & -2 & -1 & -1 & -3 \\
 \hline 
 6-5 & 3 & 0 & 0 & 3 \\
 \hline 
\end{tabular}
\caption{Table of the identified point differences between resonator pairs in the four regions of interest, indicated by the shaded columns in the top panel of Fig. \ref{fig:temp_fit}. If the first resonator thermalized faster (slower) than the other in the pair, the indicated values should all be markedly negative (positive). R1 and R4 are on SiN membranes, while the remaining resonators are on-substrate.}
\label{table:res_therm}
\end{table}
\endgroup

\section{Frequency Stability}\label{app:Freq_Stability_Comparison}
To explore any differences in the temporal frequency stability of the resonators, a single point from each was monitored over approximately 90 minutes using a VNA at $\bar{n} \sim 10$. A single point was chosen instead of a full scan over the resonator to achieve a faster repetition rate. Using initial scans across all resonators, by fitting the resonator's phase versus frequency in an interval close to resonance, a linear relationship can be extracted to convert between the two \cite{Grunhaupt_grAl_qp}. This linear relationship is formally expressed by the Barkhausen relationship, $\frac{\delta f}{f} = \frac{\delta\theta}{2 Q}$ \cite{Rubiola_book, Niepce_thesis}.

\begin{figure}
    \centering
    \includegraphics[scale=1]{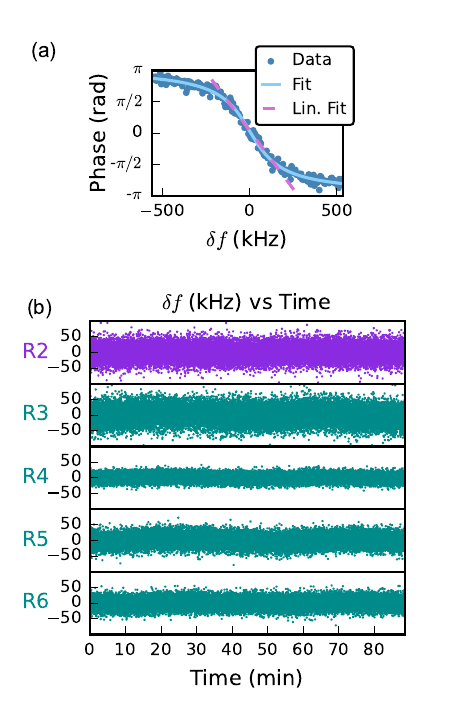}
    \caption{Frequency stability versus time. (a) Resonator phase response versus frequency. In the regime around resonance, a linear relationship between phase and frequency can be made to convert between the two. (b) Resonator frequency change versus time, converted from phase using relationships as in (a). R2 is on a SiN membrane, while R3-R6 are on-substrate.}
    \label{fig:stability}
\end{figure}

This calibration and measurement procedure was performed for on-substrate and membrane resonators on the same chip, as shown in Fig. \ref{fig:stability}. In contrast to results presented earlier, these measurements were performed in a dilution refrigerator at a temperature of 27 mK. 
To compare resonator stability, we calculated the standard deviations ($\sigma$) of the extracted frequency values over the duration of the scans \cite{Niepce_motional_narrowing}, with results of each resonator presented in Table \ref{table:res_stability}. Although certain resonators were more stable than others, no distinction was observed between membrane versus on-substrate resonators. Therefore, these results suggest that fabricating resonators on thin SiN membranes does not adversely affect their temporal frequency stability.

\begingroup
\setlength{\tabcolsep}{6pt}
\renewcommand{\arraystretch}{1.2}
\begin{table}[h]
\centering
\begin{tabular}{| c | c | c | c | c | c | c |}
 \hline
 Metric & R2 & R3 & R4 & R5 & R6 \\
 \hline
 $\sigma$ (kHz) & 21 & 24 & 10 & 17 & 15 \\
 \hline
\end{tabular}
\caption{Table of the standard deviations ($\sigma$) of the resonator frequencies over time. R2 is on membrane, while R3-R6 are on-substrate.}
\label{table:res_stability}
\end{table}
\endgroup

\section{HFSS Simulations}\label{app:HFSS_sim}
While the measured $Q_i$ values of the membrane resonators confirm that high quality superconducting circuit devices are compatible with this membrane technique, designs for integrating qubits must be verified before applying this technique to them. While simulations related to designing the chips used in this work focused on resonator frequencies and external coupling values, a greater number of parameters must be verified when extending the technique to qubits. A main concern is balancing how the etching removes most of the dielectric while certain capacitance values must still be achieved. 

\begin{figure}[h]
    \centering
    \includegraphics[scale=1]{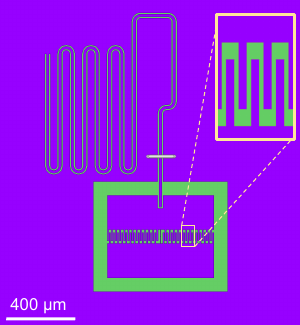}
    \caption{Design used in HFSS simulations, attaining desired capacitance and coupling values. A closer depiction of the fingers in the qubit paddles is shown in the inset.}
    \label{fig:HFSS}
\end{figure}

To address this concern and verify the compatibility between qubits and these membranes, we conducted simulations using Ansys HFSS, a commonly used software for simulating superconducting circuit designs. We found that adding ``fingers" to the qubit capacitor paddles yielded the desired capacitance ($\sim$$80 - 100$ fF); furthermore, moving the resonator coupling section closer to the qubit paddle and within the gap between it and the ground plane also yielded the desired qubit-resonator coupling ($\sim$$100$ MHz), both values typical of other transmon designs \cite{Nguyen_bosonic_ladder}. Both of these simulations were performed using the Eigenmode solver of HFSS. To identify the qubit capacitance, a lumped element inductor was placed between the capacitor paddles, modelled as 2D sheets with a perfect E boundary condition. From the simulated frequency of the associated mode, the corresponding capacitance was calculated. Meanwhile, $g$ was simulated by sweeping the inductance of the qubit mode through the resonator mode and scaling the avoided crossing difference back to the frequencies of interest. With these simulations as guidance, one can begin to explore placing qubits on SiN membranes.

\newpage
\section*{References}
\bibliography{biblio}

\end{document}